\documentclass[onecolumn,12pt]{aastex631}
\usepackage{graphicx,subfigure,amsmath,amsfonts,amssymb,multirow,extarrows}
\usepackage{bm}

\def\aap{Astron.\ Astrophys.\ }

\def\apjl{Astrophys.\ J.\ Lett.\ }
\def\apjs{Astrophys.\ J.\ Supp.\ }

\newcommand{\onerho}{$p(\rho_{\rm sat})~[{\rm 10^{33}~dyn/cm^2}]$}
\newcommand{\tworho}{$p(2\rho_{\rm sat})~[{\rm 10^{34}~dyn/cm^2}]$}
\newcommand{\sixrho}{$p(6\rho_{\rm sat})~[{\rm 10^{35}~dyn/cm^2}]$}
\newcommand{\mmax}{$M_{\rm max}$ $[M_\odot]$}
\newcommand{\ronefour}{$R_{1.4}$ $[{\rm km}]$}
\newcommand{\rtwo}{$R_{2}$ $[{\rm km}]$}
\newcommand{\lonefour}{$\Lambda_{1.4}$}

\begin{document}
\newcommand{\PMO}{Key Laboratory of Dark Matter and Space Astronomy, Purple Mountain Observatory, Chinese Academy of Sciences, Nanjing, 210033, People's Republic of China.}
\newcommand{\USTC}{School of Astronomy and Space Science, University of Science and Technology of China, Hefei, Anhui 230026, People's Republic of China.}
\newcommand{\NJNU}{Department of Physics and Institute of Theoretical Physics, Nanjing Normal University, Nanjing 210046, People's Republic of China.}

\title{Bayesian nonparametric inference of neutron star equation of state via neural network}
\author[0000-0001-9034-0866]{Ming-Zhe Han}
\author[0000-0002-9078-7825]{Jin-Liang Jiang}
\author[0000-0001-9120-7733]{Shao-Peng Tang}
\author[0000-0002-8966-6911]{Yi-Zhong Fan}
\email{Corresponding author: yzfan@pmo.ac.cn}
\affiliation{\PMO}
\affiliation{\USTC}
\date{\today}

\begin{abstract}
We develop a new nonparametric method to reconstruct the Equation of State (EoS) of Neutron Star with multimessenger data. As an universal function approximator, the Feed-Forward Neural Network (FFNN) with one hidden layer and a sigmoidal activation function can approximately fit any continuous function. Thus we are able to implement the nonparametric FFNN representation of the EoSs. This new representation is validated by its capabilities of fitting the theoretical EoSs and recovering the injected parameters. Then we adopt this nonparametric method to analyze the real data, including mass-tidal deformability measurement from the Binary Neutron Star (BNS) merger Gravitational Wave (GW) event GW170817 and mass-radius measurement of PSR J0030+0451 by {\it NICER}. We take the publicly available samples to construct the likelihood and use the nested sampling to obtain the posteriors of the parameters of FFNN according to the Bayesian theorem, which in turn can be translated to the posteriors of EoS parameters. Combining all these data, for a canonical 1.4 $M_\odot$ neutron star, we get the radius $R_{1.4}=11.83^{+1.25}_{-1.08}$ km and the tidal deformability $\Lambda_{1.4} = 323^{+334}_{-165}$ (90\% confidence interval).Furthermore, we find that in the high density region ($\geq 3\rho_{\rm sat}$), the 90\% lower limits of the $c_{\rm s}^2/c^2$ ($c_{\rm s}$ is the sound speed and $c$ is the velocity of light in the vacuum) are above $1/3$, which means that the so-called conformal limit (i.e., $c_{\rm s}^2/c^2<1/3$) is not always valid in the neutron stars.
\end{abstract}


\section{Introduction}
Neutron stars (NSs) are natural laboratories to study the properties of the dense matter (see Refs.\citep{2012ARNPS..62..485L, 2016PhR...621..127L, 2016ARA&A..54..401O, 2017RvMP...89a5007O} for reviews), in which the extreme conditions (i.e., the conditions in the NS core) are hard to achieve in terrestrial experiments.
Usually, the matter in NS (in most of its lifetime except for the birth and death/merger) can be well described by the zero-temperature pressure-density relation, i.e., the equation of state (EoS). Up to now, the properties of the low density matter have been well constrained by the nuclear theories/experiments \citep{2013ApJ...771...51L,2019JPhG...46g4001K,2018PhRvL.121f2701L}, and the behavior of super high density matter is robustly predicted by the perturbative quantum chromodynamics (pQCD) theory \citep{2018RPPh...81e6902B}.
However, the matter in the transition area (between the low and the super high density) is not clear yet. For instance, the presence of a hadron-quark phase transition \citep{2021PhRvD.103f3026T} or not is quite uncertain.
The observations of NSs are essential to reveal the properties of the matter in the medium density region. 

On August 17, 2017, the gravitational wave (GW) signal from the coalescence of binary neutron star (BNS) was detected by the LIGO/Virgo detectors \citep{2017PhRvL.119p1101A,2019PhRvX...9a1001A}, which provides new avenues to constrain the EoS via the NS matter effects.
Benefiting from this remarkable GW event (GW170817), various studies have been implemented to better understand the EoS of NSs \citep{2018A&A...616A.105S, 2018PhRvL.120z1103M, 2018PhRvL.121i1102D, 2018PhRvL.121p1101A, 2019ApJ...887L..22R, 2019ApJ...887L..24M, 2019PhRvD..99h4049L, 2019PhRvD..99j3009M, 2020ApJ...888...12M, 2020ApJ...888...45T, 2020ApJ...893L..21R, 2020PhRvC.102e5801K, 2020PhRvD.101f3007E, 2020PhRvD.101l3007L, 2020arXiv200203210Z, 2020arXiv200801582B, 2021PhRvC.103b5803H, 2021PhRvL.126f1101A}.
With the improvement on the sensitivities of GW detectors, the detectable volume will be significantly enhanced, and an increasing sample of BNS mergers will be available to constrain the EoS.
Nevertheless, in this work we do not take into account GW190425 \citep{2020ApJ...892L...3A} because of its weak tidal effect (due to the low signal to noise ratio and massive components).
Moreover, its possible black hole-neutron star merger nature \citep{2020ApJ...891L...5H} may influence the inference.
In addition to the GW signals, the mass$-$radius ($M-R$) measurements of the isolated NS PSR J0030+0451 were recently reported by the {\it NICER} collaboration \citep{2016SPIE.9905E..1HG}.
These measurements \citep{2019ApJ...887L..21R, 2019ApJ...887L..24M} were based on the pulse profile modeling of the X-ray emission from the hot spots on the NS surface, which are more reliable and accurate than traditional/indirect spectroscopic measurements/estimates.
Such data are valuable to constrain the NS EoS \citep{2019ApJ...887L..24M, 2019ApJ...887L..22R, 2018A&A...616A.105S}. 
Tighter constraints on the EoS have been reported in the joint analysis of these multimessenger observations of NSs \citep{2018A&A...616A.105S, 2019ApJ...887L..22R, 2019ApJ...887L..24M, 2020ApJ...892...55J, 2020ApJ...893L..21R, 2020PhRvD.101l3007L, 2020arXiv200203210Z, 2020arXiv200801582B, 2021PhRvD.103f3026T, 2021PhRvC.103b5803H, 2021PhRvL.126f1101A} under the assumption that all NSs share the same EoS.

The phenomenological parameterization methods, including for instance the spectral representation \citep{2010PhRvD..82j3011L} and the piecewise polytropic expansion \citep{2009PhRvD..79l4032R, 2016ApJ...820...28O, 2017ApJ...844..156R}, have been widely adopted to study the EoS of NSs.
These parametric models are described as $\mathcal{F} = \{f_{\rm \bm{\theta}} : \bm{\theta} \in \bm{\Theta} \subset \mathbb{R}^p\}$, where $f_{\rm \bm{\theta}}$ is the probability density function (PDF) of the underlying probability distribution as a function of $\bm{\theta}$.
The inference aims to get the credible intervals of $\bm{\theta}$. However, a specific parametric form may be model-dependent and the range and type of inferences are likely limited, too.
In order to avoid model mis-specification, the extension beyond the parametric assumptions may be needed to attain more robust models.
In this situation, one can consider a nonparametric model where the functional space is so large that the approach will not depend on a specific parametric form.
Such nonparametric approaches are considered as the models with infinite dimensional parameter.
However, in practice we usually use finite dimensional parameter instead, since it can provide an approximation to the infinite sum.
The nonparametric model via Gaussian process (GP) has been implemented to the studies of the EoS \citep{2019PhRvD..99h4049L, 2020PhRvD.101f3007E, 2020PhRvD.101l3007L}.
These works use GP and Bayesian methods to infer the NS EoS from multimessager data.
The posteriors are calculated by using the Monte Carlo integration, which is relatively inefficient in comparison to the Markov-chain Monte Carlo (MCMC) method.
The MCMC approach has not been adopted in these previous works because the associated jump proposals are thought to be nontrivial.
In practice, one can use a finite dimensional EoS table to represent the real EoS, then the EoS model has a finite number of random variables (RVs) at the discrete set of densities drawn from the GP, whose covariance matrix is determined by a set of hyper parameters, and the hyper parameters is chosen based on the training set (i.e., the theoretical EoSs).
However, to get a precise approximation one often take $\gtrsim$ 100 dimensions, thus the model will have $\gtrsim$ 100 parameters, and it is extremely expensive in computation to perform the Bayesian inference, while as one will see below the number of RVs of the model in this work is much fewer.
Besides, the GP EoS prior relies on the choice of the training set, therefore, it might be limited by the current knowledge of EoS \citep[see][for more details]{2019PhRvD..99h4049L,2020PhRvD.101f3007E}.
\citet{2018PhRvD..98b3019F,2020PhRvD.101e4016F,2021arXiv210108156F} have studied the methodology of the machine learning method for EoS.
They take the use of the Feed-Forward Neural Network (FFNN) to map the neutron star data to the EoS parameters, while in their approach the training data are still generated by a specific parametric form.
Different from their work, we do not consider the FFNN as an interpolation tool between the observations and the EoS, but as a representation of the EoS itself, then we can combine it with the Bayesian statistical framework, which is more explainable.

In this work we use the Bayesian nonparametric method via the FFNN to reconstruct the NS EoS with multimessenger data.
We describe in details the nonparametric representation of NS EoS along with fitting the theoretical EoSs in Sec.\ref{nonp method}, and introduce the observed data used for the Bayesian inference in Sec.\ref{bayes}.
The main results of our work is presented in Sec.\ref{results}, including the simulated/real data. Finally, Sec.\ref{summary} is our summary and conclusion.

\section{Method}
\subsection{Nonparametric representation of EoS}
\label{nonp method}
Consider a general regression problem of fitting an output $y$ as a function of input $x$: 
\begin{equation}
    y = f(x) + \epsilon, 
\end{equation}
where the $f$ is called the mean function and the $\epsilon$ is called the residual, usually they are independent.
Based on which part of the model is nonparametric, the nonparametric models can be divided into three types: nonparametric mean function, nonparametric residual distribution, and fully nonparametric regression.
Furthermore, the nonparametric mean function methods consist of basis expansion, b-spline, Gaussian process, and the regression tree \citep[for more details see Chapter 4 of][]{muller_Bayesian_2015}.
In this work we use the nonparametric mean function method via the FFNN.
An FFNN with one hidden layer can be written as 
\begin{equation}
    y = \sum^M_{\rm i=1} w_{\rm 2i} \sigma (w_{\rm 1i} x + b_{\rm i}) + \epsilon, 
\label{eq:FFNN}
\end{equation}
where $y$ is the dependent variable, $x$ is the independent variable, $M$ is the number of neural nodes, $w_{\rm 1i}$ and $w_{\rm 2i}$ are the weights, $b_{\rm i}$ is the bias, $\epsilon$ is the residual, and $\sigma$ is the activation function (a nonlinear function).
\citet{Cybenko1989} and others pointed out that if we use a sigmoidal function as the activation function, the finite sums in the form of Eq.(\ref{eq:FFNN}) are dense in the functional space of continuous functions defined in the n-dimensional unit cube $[0, 1]^n$ (for more details see \citet{Cybenko1989}), which means that we can use Eq.(\ref{eq:FFNN}) with a sigmoidal activation function and finite $M$ to fit any real continuous functions.
This property exactly matches the feature of nonparametric models (i.e., the modeling should be flexible and robust) we have mentioned above. 
It thus gives us the possibility to take it as a nonparametric representation of the EoS.
In this work we use the sigmoid function as the activation function
\begin{equation}
    \sigma(x) = \frac{1}{1 + e^{-x}},
\end{equation}
which yields $\sigma(x) \rightarrow 0(1)$ when $x \rightarrow -\infty(+\infty)$.
Both the input layer and the output layer have one node without activation function, as for the hidden layer, we have tried several settings (5, 10, 20, and more).
We find that if the number of nodes is less than 10, the model could not represent all the EoSs with relatively small errors, and as the number of nodes increase, the efficiency of the sampling algorithm decreases.
Therefore, we choose 10 nodes for this work.
Besides, it should be noted that we call our model as nonparametric model for its ability of representing almost all theoretical EoSs (as shown below), even though it only has 31 parameters which is far away from infinity.

In the zero temperature case which we consider here, the NS EoS can be represented by the relation between the total energy density $\varepsilon$ and the pressure $p$. Thus in order to constrain the EoS, we need to reconstruct the $\varepsilon(p)$ relation from the data.
However, not all the $\varepsilon(p)$ function can be a realistic EoS, unless it satisfies two conditions:
\begin{itemize}
    \item [1)]
    The microscopical stability condition, i.e., $\frac{dp}{d\varepsilon} \geq 0.$
    \item [2)]
    The causality condition, i.e., the sound speed $c_{\rm s} = \sqrt{\frac{dp}{d\varepsilon}} < c$, where $c$ is the speed of light in vacuum.
\end{itemize}
Fortunately, it has been shown that the auxiliary variable
\begin{equation}
\label{eq:phi}
    \phi = \log(c^2 \frac{d\varepsilon}{dp} - 1)
\end{equation}
can automatically satisfy these two conditions \citep{2010PhRvD..82j3011L}, thus it is widely used in parameterizing the EoS \citep{2019PhRvD..99h4049L,2020PhRvD.101f3007E,2020PhRvD.101l3007L}.
We therefore take it as the output variable, and the $\log(p)$ as the input variable of the FFNN.
We standardize the inputs of the FFNN before the inference, i.e., $x_{\rm std}=(x-\mu)/\sigma$, where $\mu$ and $\sigma$ are the mean and standard deviation of $x$, respectively.
In the low density ($\sim 0.3\rho_{\rm sat}$) range we match the constructed EoS with the known EoS SLy \citep{2001A&A...380..151D}\footnote{In practice, when integrating the macroscopic properties of neutron star, we divide the EoS constructed by the FFNN into a finite dimensional (100) EoS table (The pressures of these nodes are logarithmically uniform in $[p_{\rm SLy}(\sim 0.3 \rho_{\rm sat}), 5 \times 10^{15}~{\rm g/cm^3}~\cdot~c^2]$, and by using Eq. \ref{eq:phi} we can obtain the $\varepsilon$ via fixed step-size integration.), and between adjacent discrete density points we apply the method described in Appendix B of \citet{2014PhRvD..89f4003L} to interpolate the EoS tables.
The Eq. (\ref{eq:phi}) gives an one to one map between the auxiliary variable $\phi$ and the first derivative of energy density with respect to the pressure.
If the initial condition of pressure and energy density is further given, then this would yield a unique EoS.
So at the matching point, we fix the first point of the constructed EoS by the corresponding pressure and energy density of the EoS SLy.}.
Additionally, according to the theoretical conjecture that the energy of the unitary gas is less than the energy of pure neutron matter and the various experimental constraints on the symmetry energy parameters\citep{2017ApJ...848..105T,2014EPJA...50...40L}, we constrain the $p(\rho_{\rm sat})$ with $3.12 \times 10^{33}~{\rm dyn/cm^2} \leqslant p(\rho_{\rm sat}) \leqslant 4.70 \times 10^{33}~{\rm dyn/cm^2}$ \citep{2019ApJ...885...39J}, where $\rho_{\rm sat} = 2.7 \times 10^{14}~{\rm g/cm^3}$ is the saturation density.
And the pressure $p(1.85\rho_{\rm sat})$ is limited to $p(1.85\rho_{\rm sat}) \geqslant 1.21 \times 10^{34}~{\rm dyn/cm^2}$ according to the theoretical conjecture that the two nucleon interaction pressure is an absolute lower bound because the three-body interactions in pure neutron matter are always repulsive \citep{2016ApJ...820...28O}.
EoSs violating these conditions will be directly rejected in the sampling procedure.

\begin{figure}
    \centering
    \subfigure{\label{fig:fit phi}
        \includegraphics[width=0.49\columnwidth]{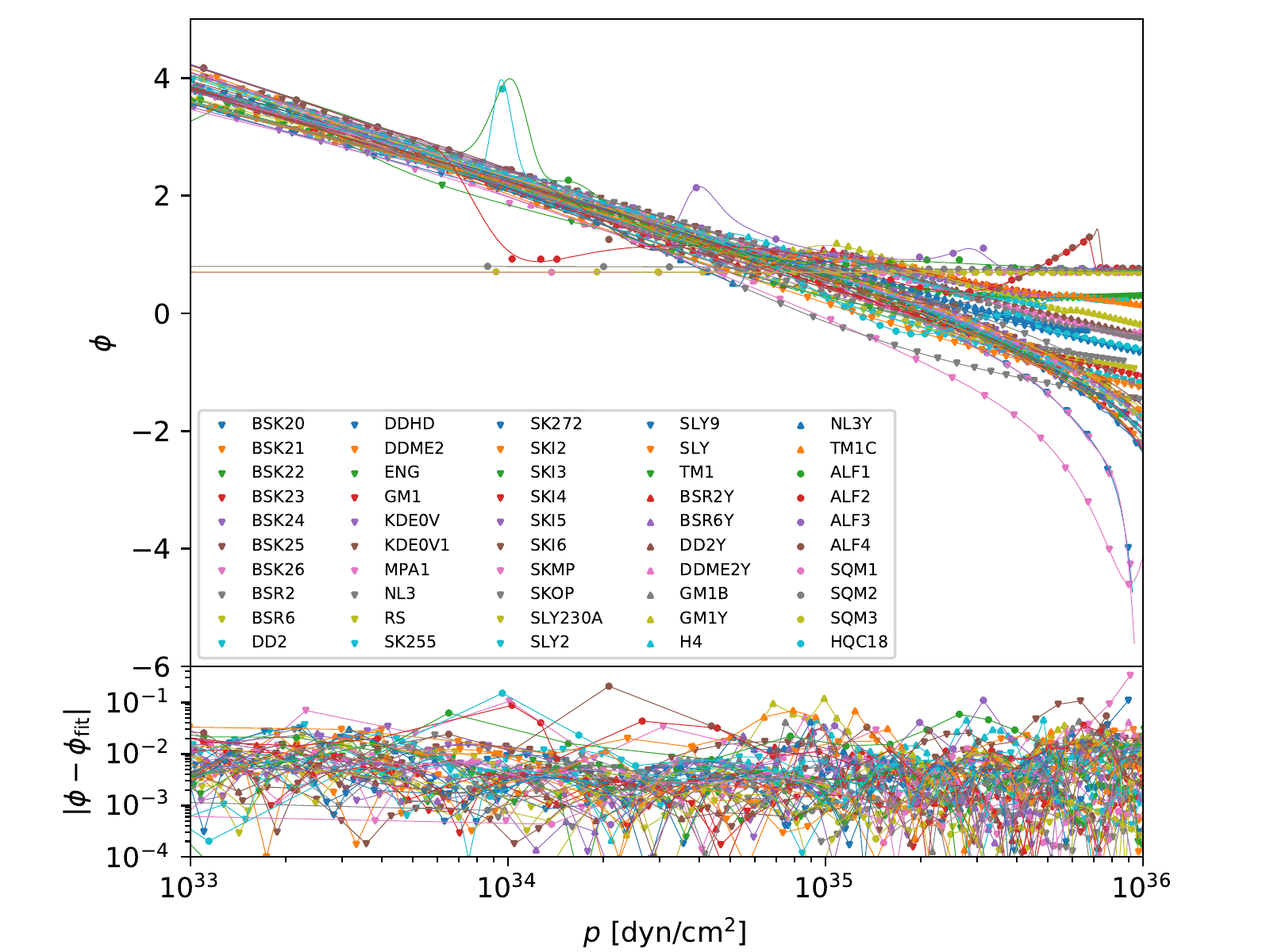}}
    \subfigure{\label{fig:fit gamma}
        \includegraphics[width=0.49\columnwidth]{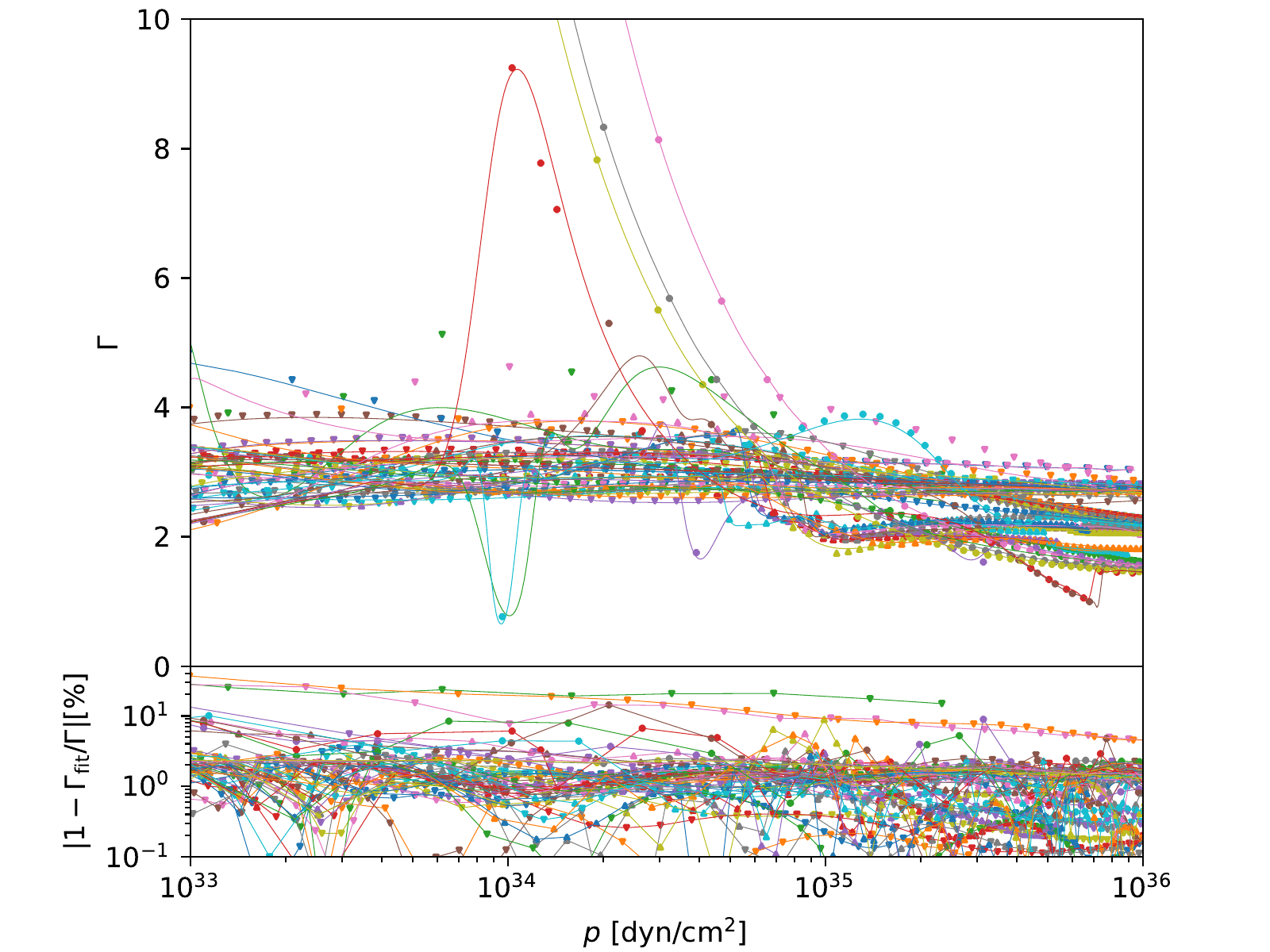}}
    \caption{Fitting results of the $\phi(p)$ (left panel) and the corresponding $\Gamma(p)$ (right panel) for theoretical EoSs. In the top plots of each panel, the scatters denote the theoretical values, while the solid lines are the fitting values by the FFNN. In the bottom plots, the absolute errors for the $\phi(p)$ fitting and the relative error for the corresponding $\Gamma(p)$ are displayed. The EoSs we fitted are divided into three distinct groups, hadronic (down triangle), hyperonic (up triangle), and quarkyonic (circle).}
    \label{fig:fit}
    \hfill
\end{figure}

To illustrate the accuracy and efficiency of this new nonparametric representation, we first implement the FFNN to fit the theoretical EoSs.
We use the Python library Keras \citep{2018ascl.soft06022C} with TensorFlow \citep{2016arXiv160508695A} as a backend. We take the mean square errors as the loss function, i.e., 
\begin{equation}
    L_{\rm mse} = \frac1n \sum^N_{i=1} (\hat{y_i}-y_i)^2,
\end{equation}
where the $\hat{y_i}$ is the true value and the $y_i$ is the predicted value.
As for the optimization method we use Adam \citep{2014arXiv1412.6980K}.
Three groups of EoSs that we fitted are characterized by the different compositions (see Fig.\ref{fig:fit}):
\begin{itemize}
    \item [1)]
    Hadronic EoSs, i.e., BSK20, BSK21, BSK22, BSK23, BSK24, BSK25, BSK26 \citep{2010PhRvC..82c5804G,2013PhRvC..88b4308G};
        BSR2, BSR6 \citep{2010PhRvC..81c4323A};
        DD2, DDHD, DDME2 \citep{2014ApJS..214...22B,2004NuPhA.732...24G,2005PhRvC..71b4312L};
        ENG \citep{1996ApJ...469..794E};
        GM1 \citep{1991PhRvL..67.2414G};
        KDE0V, KDE0V1 \citep{2015PhRvC..92e5803G,2005PhRvC..72a4310A};
        MPA1 \citep{1987PhLB..199..469M};
        NL3 \citep{1997PhRvC..55..540L};
        RS \citep{1986PhRvC..33..335F};
        SK255, SK272, SKI2, SKI3, SKI4, SKI5, SKI6, SKMP, SKOP \citep{2003PhRvC..68c1304A,1995NuPhA.584..467R,1996PhRvC..53..740N,1989PhRvC..40.2834B,1999PhRvC..60a4316R};
        SLY230A, SLY2, SLY9, SLY \citep{1997NuPhA.627..710C,1995eiei.book.....C,2001A&A...380..151D};
        TM1 \citep{1994NuPhA.579..557S}.
    \item [2)]
    Hyperonic EoSs, i.e., BSR2Y, BSR6Y \citep{2016PhRvC..94c5804F};
        DD2Y, DDME2Y \citep{2016PhRvC..94c5804F};
        GM1B, GM1Y \citep{2016PhRvC..94c5804F,2014MNRAS.439..318G};
        H4 \citep{2006PhRvD..73b4021L};
        NL3Y \citep{2016PhRvC..94c5804F};
        TM1C \citep{2014MNRAS.439..318G}.
    \item [3)]
    Quark matter or hybrid EoSs, i.e., ALF1, ALF2, ALF3, ALF4 \citep{2005ApJ...629..969A};
        SQM1, SQM2, SQM3 \citep{1995PhRvD..52..661P};
        HQC18 \citep{2019ApJ...885...42B}.
\end{itemize}
The direct variable we fitted is the $\phi(p)$, we calculate the absolute errors in fitting $\phi(p)$, i.e., $|\phi - \phi_{\rm fit}|$, and the values are as low as $\sim 10^{-4} - 10^{-1}$, while the absolute value of $\phi$ is $\sim$ $0.1-10$.
Furthermore, we use the fitted $\phi(p)$ to calculate the corresponding $\Gamma(p)$, which is defined as $\Gamma(p) = [(\varepsilon+p)/p]dp/d\varepsilon$, and the relative errors, i.e., $|1-\Gamma_{\rm fit}/\Gamma|$, are as low as $\sim$ 0.1\% $-$ 10\%.
Thus it is reasonable to conclude that the FFNN can well represent the various realistic EoSs.

\subsection{Bayesian Inference}\label{bayes}
After constructing the nonparametric representation of EoS with FFNN, we can optimize the parameters (i.e., the weights, bias, and residual of the FFNN) with the observation data.
The standard approach for neural networks is to approximate a minimal loss.
From a statistician point of view, we can consider it as MAP (Maximum A Posteriori) or MLE (Maximum Likelihood Estimation) by constructing a proper likelihood function.
These two point estimate methods are relatively easy to realize, but can not give us the credible intervals of the parameters that we are interested in.
Then we use the Bayesian statistical framework to obtain the posterior distributions of the parameters.

Assuming that all neutron stars share the same EoS, the likelihood $\mathcal{L}$ for Bayesian inference can be written as
\begin{equation}
    \mathcal{L}=\mathcal{L}_{\rm GW}(d\mid\vec{\theta}_{\rm GW}, \vec{\theta}_{\rm EoS}) \times \mathcal{L}_{\rm Nuc}(\vec{\theta}_{\rm EoS}) \times \mathcal{L}_{\rm M_{\rm max}}(\vec{\theta}_{\rm EoS}) \times \mathcal{P}(M(\vec{\theta}_{\rm EoS}, h_{\rm c}), R(\vec{\theta}_{\rm EoS}, h_{\rm c})\,),
    \label{eq:likelihood}
\end{equation}
where $M$, $R$, and $h_{\rm c}$ are mass, radius, and pseudo enthalpy of the star's core, respectively.
The $\vec{\theta}_{\rm GW}$ and $\vec{\theta}_{\rm EoS}$ are respectively the parameters for GW and FFNN.
$\mathcal{L}_{\rm GW} = \mathcal{P}(m_1, m_2, \Lambda_1(m_1, \vec{\theta}_{\rm EoS}), \Lambda_2(m_2, \vec{\theta}_{\rm EoS})\mid d)$ is the interpolated GW likelihood that marginalized over the other nuisance parameters \citep{2020MNRAS.499.5972H}, $d$ is the strain data of GW170817, $m_{\rm 1, 2}$ and $\Lambda_{\rm 1, 2}$ are primary/secondary mass and its tidal deformability.
$\mathcal{L}_{\rm Nuc}(\vec{\theta}_{\rm EoS})$ takes $1$ only if all the nuclear constraints are satisfied, otherwise it takes $0$ (In practice, we set the log(0) to be $-10^{100}$).
For the $\mathcal{L}_{\rm M_{\rm max}}(\vec{\theta}_{\rm EoS})$ we follow the method in Sec. 3.1.2 of \citet{2020ApJ...888...12M}, that is
\begin{equation}
\mathcal{L}_{\rm M_{\rm max}}(\vec{\theta}_{\rm EoS}) = \int^{M_{\rm max}(\vec{\theta}_{\rm EoS})}_{0} P(M) dM,
\end{equation}
where the $P(M)$ is the probability distribution for the mass of PSR J0740+6620 \citep{2021arXiv210400880F}.
While for constructing the $\mathcal{P}(M, R)$, we use the Gaussian kernel density estimation (KDE) with the $M, R$ samples of PSR J0030+0451 from {\it NICER} \citep{riley_thomas_e_2019_3386449, miller_m_c_2019_3473466}.

We implement the enthalpy-based formulas of \citet{2014PhRvD..89f4003L} to solve the Tolman-Oppenhimer-Volkoff and Regge-Wheeler equations to get the mass, radius and tidal deformability of neutron star.
The prior of $h_{\rm c}$ is based on the method described in the Sec. 5.1 in \citet{2021arXiv210506979M}. We uniformly sample $x$ in (0, 1), the central enthalpy is $h_{\rm c} = h_{\rm min} + x^2 (h_{\rm max} - h_{\rm min})$, where the $h_{\rm min}$ and $h_{\rm max}$ are respectively the central enthalpy of a 1 $M_\odot$ and a maximum-mass non-rotating neutron star for the EOS under consideration. And the quadratic prior is because the central enthalpy changes rapidly near the maximum mass. Then the $h_{\rm c}$ can be combined with $\vec{\theta}_{\rm EoS}$ to calculate a pair of macroscopic observables ($M$ and $R$), and hence the $\mathcal{P}(M, R)$. For the $\mathcal{L}_{\rm GW}$, we sample the chirp mass $\mathcal{M} = (m_1m_2)^{3/5}/(m_1+m_2)^{1/5}$ and the mass ratio $q=m_2/m_1$ ($m_1>m_2$) instead of the component masses $m_{1, 2}$ with proper uniform priors.
We can then map the transformed source frame masses $m_{\rm 1, 2}^{\rm src}$ to $\Lambda_{\rm 1, 2}$ with the EoS parameters $\vec{\theta}_{\rm EoS}$, and calculate the $\mathcal{L}_{\rm GW}$.
As for the parameters of FFNN, we set all the 31 parameters to be uniformly distributed in $(-5, 5)$, the bounds are chosen as the $5 \sigma$ credible interval of the standard normal distribution, since the inputs have been standardized.
We use the nested sampling algorithm {\sc PyMultiNest} \citep{2014A&A...564A.125B} in {\sc Bilby} \citep{2019ApJS..241...27A} to sample the posterior distributions of all those parameters, typically with 1000 live points, a sampling efficiency of 0.01 (without constant efficiency), and a tolerance of 0.5.

\section{Results}\label{results}
\subsection{Simulated data analysis}\label{sim}
To illustrate the performance of our nonparametric method on recovering the EoS from observations, we simulate mock data calculated by theoretical EoSs and perform the same procedure applied to the analysis of the real data except for the nuclear constraints.
We select three NSs' masses based on the most probable masses in the posterior samples of GW170817 \footnote{\url{https://dcc.ligo.org/LIGO-P1800061/public}} and PSR J0030+0451 \citep{riley_thomas_e_2019_3386449, miller_m_c_2019_3473466}, then use the theoretical EoS to calculate the corresponding tidal deformabilities and radius. The mock likelihood function is multidimensional Gaussian, with the means given by simulated values and the comparable variances based on current quality of the real data samples.
We choose four tabulated EoSs as the theoretical model, including three hadronic EoSs (WFF1 \citep{1988PhRvC..38.1010W}, H4, MS1 \citep{1996NuPhA.606..508M}), and a hybrid EoS (ALF2), where the hadronic EoSs range from relatively soft to stiff. 

\begin{figure}
    \centering
    \includegraphics[width=0.8\columnwidth]{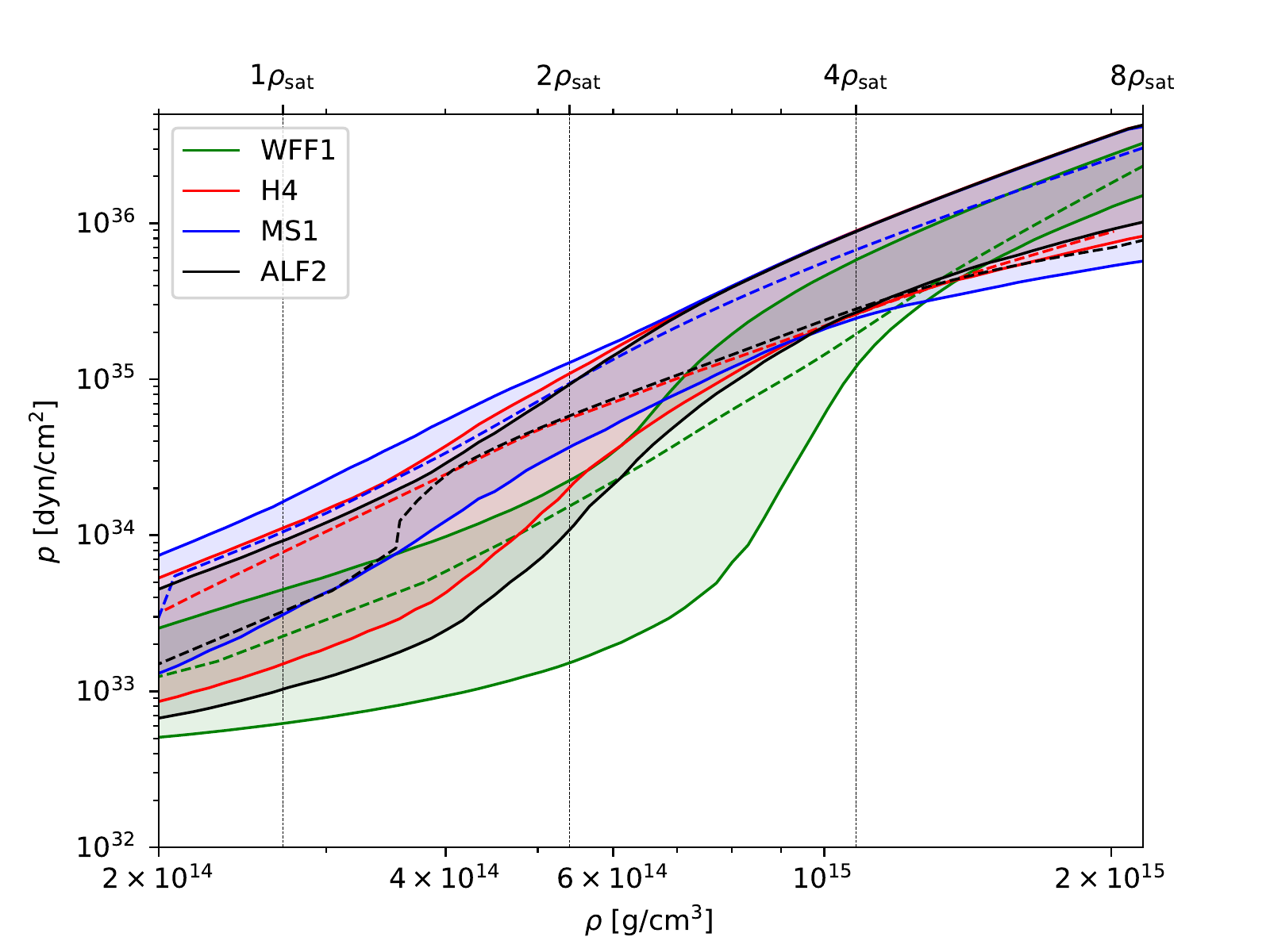}
    \caption{Reconstructed $p(\rho)$ for four injected EoSs: the case of three hypothesized neutron stars. The colored regions are the 90\% credible intervals for the pressure $p$ at corresponding rest mass density $\rho$ of each EoS, while the dashed curves denote the theoretical EoSs. The vertical dashed lines represent multiples of $\rho_{\rm sat}$.}
    \label{fig:inj_rec}
    \hfill
\end{figure}

The credible intervals of the recovered parameters are summarized in Tab.\ref{tab:inj_rec}.
We can see that there are no egregious errors in the recovering process and all the injected values lie in the 90\% credible intervals of the recovered values, which is in keeping with our expectations.
Fig.\ref{fig:inj_rec} shows the reconstructed EoSs ($p(\rho)$) of all the injected theoretical EoSs, and the posterior regions comfortably include the $\varepsilon(p)$ curves of the corresponding injected EoSs.
Therefore, it is rather appropriate to use this nonparametric model to reconstruct the EoS with the observational data.

\begin{table*}[]
    \caption{Recovered values versus injected ones for $\Lambda_1$, $\Lambda_2$, and $R$. The intervals are the recovered 90 \% credible intervals, and the corresponding injected values are denoted in round brackets. These injected values of $\Lambda_1$ and $\Lambda_2$ are selected based on the most probable masses of the GW170817 and mapped through the theoretical EoSs, while the injected values of $R$ are mapped the same way through the most probable mass of {PSR J0030+0451}.}
    \begin{center}
        \begin{tabular}{lcccccc}
            \hline \hline
            EoS & $\Lambda_1$ & $\Lambda_2$ & $R~[{\rm km}]$ \\
            \hline
            WFF1 & $117^{+117}_{-61}~(106)~$ & $296^{+249}_{-151}~(294)~$ & $10.21^{+1.18}_{-1.11}~(10.42)~$ \\
            H4 & $483^{+544}_{-288}~(630)~$ & $1175^{+1404}_{-665}~(1672)~$ & $13.24^{+1.72}_{-1.82}~(13.69)~$ \\
            MS1 & $839^{+1133}_{-539}~(1181)~$ & $2164^{+2598}_{-1385}~(2904)~$ & $14.66^{+1.84}_{-2.17}~(15.06)~$ \\
            ALF2 & $376^{+421}_{-220}~(550)~$ & $928^{+1028}_{-523}~(1339)~$ & $12.61^{+1.68}_{-1.69}~(13.15)~$ \\
            \hline \hline
        \end{tabular}
    \end{center}
    \label{tab:inj_rec}
\end{table*}

\subsection{Real data analysis}

Our direct results are the posterior distributions of the weights and bias parameters of the hidden layer, but these parameters don't have a strong correlation with the output variable (this is why we call it hidden layer). We thus transform them to the distribution of the output variable $\phi$, and further calculate the parameters of the EoS that we are interested in.
The results can be divided into five groups:

\begin{itemize}
    \item [1)]
    The microscopical stability condition, the causality condition, and the request of $M_{\rm max} \in (1.4, 3)~M_\odot$ (i.e., Prior).
    \item [2)]
    The constraints of 1) and the mass information of PSR J0740+6620 (i.e., PSR).
    \item [3)]
    The constraints of 2) and the nuclear constraints, $3.12 \times 10^{33}~{\rm dyn/cm^2} \leqslant p(\rho_{\rm sat}) \leqslant 4.70 \times 10^{33}~{\rm dyn/cm^2}$ \citep{2019ApJ...885...39J} and $p(1.85\rho_{\rm sat}) \geqslant 1.21 \times 10^{34}~{\rm dyn/cm^2}$ \citep{2016ApJ...820...28O} (i.e., PSR+Nuclear).
    \item [4)]
    The constraints of 3) and the data of GW170817 (i.e., PSR+Nuclear+GW).
    \item [5)]
    The constraints of 4) and the data of PSR J0030+0451 (i.e., PSR+Nuclear+GW+NICER).
\end{itemize}

\begin{figure}
\centering
\includegraphics[width=\columnwidth]{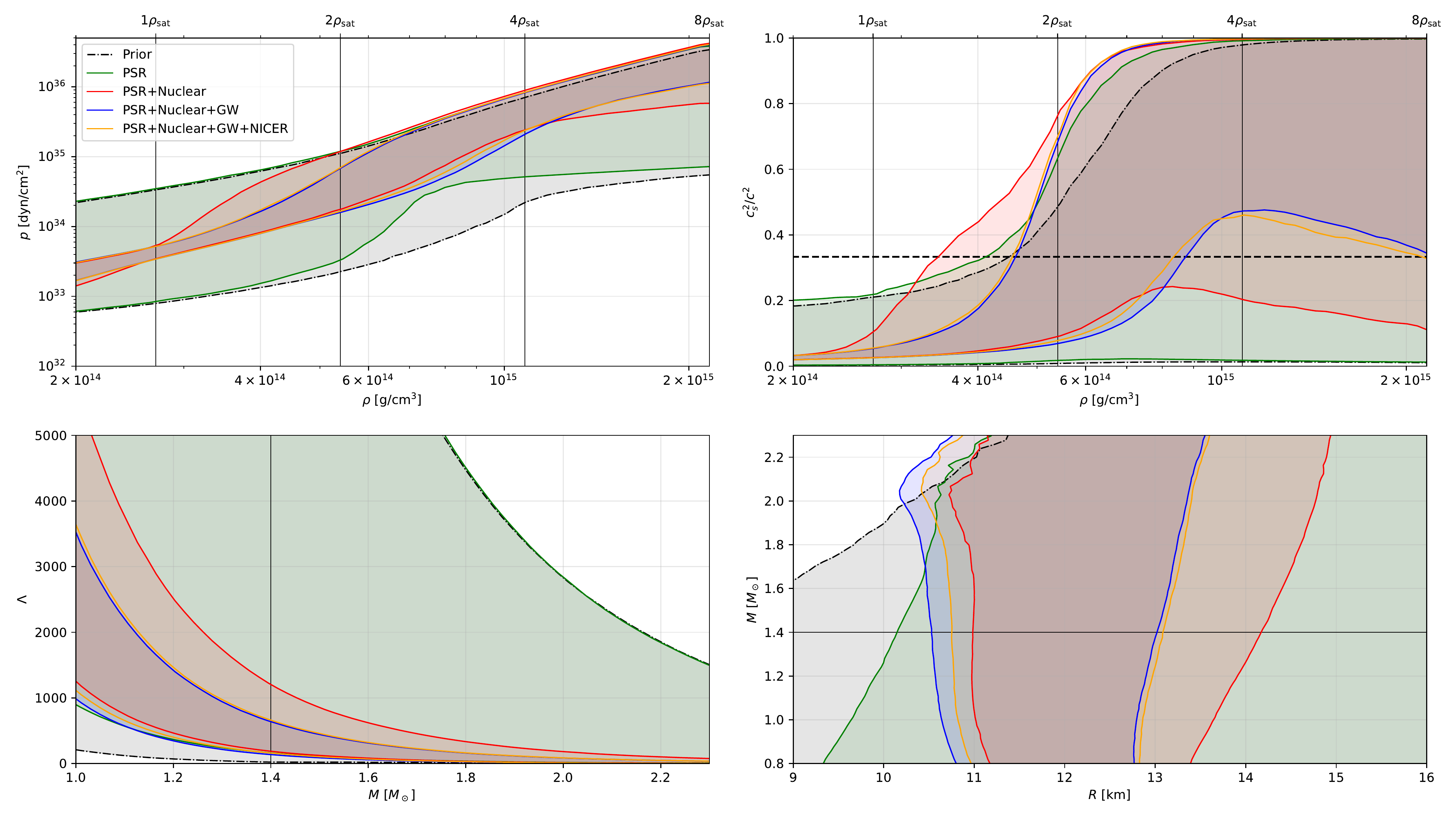}
\caption{The 90\% symmetric credible intervals for the pressure (upper left) as a function of rest-mass density, the squared speed of sound (upper right) as a function of rest-mass density, the tidal deformability (lower left) as a function of mass, and the radius (lower right) as a function of mass. To obtain these credible intervals, first a series of curves are calculated, e.g., the mass-radius curves that can be obtained with EoS parameters by uniformly spacing the masses from $(0.8~M_\odot, M_{\rm max})$, and each curve can be linearly interpolated, then for a given mass or density, the symmetric credible interval can be calculated throughout all of the EoS parameters. Notice that for an EoS does not support the given mass or density, we will discard it in the calculation of credible intervals. Black contours are the prior ranges, the green contours denote the results when we add the mass information of PSR J0740+6620, and the red contours are the results considering the nuclear constraints. The posteriors of the GW170817 and combined GW170817/PSR J0030+0451 likelihood are respectively drawn in blue and orange contours. The black vertical lines in the two upper panels stand for the $\rho_{\rm sat}$, $2\rho_{\rm sat}$, $4\rho_{\rm sat}$, and $8\rho_{\rm sat}$, correspondingly. The horizon dashed line in upper right panel represents the conformal limit, i.e., $c_{\rm s}^2/c^2<1/3$, while the vertical (horizon) solid line in lower left (right) panel denotes the mass of $1.4~M_\odot$.}
\label{fig:posterior}
\hfill
\end{figure}

The mass information of the PSR J0740+6620 raises the pressure at $\sim 3\rho_{\rm sat}$, while the possible region of the $\rho-p$ relation has been effectively narrowed down by the nuclear constraints (see upper left panel of Fig.\ref{fig:posterior}).
The results yielded by adding the GW data indicate that stiffening at $\sim 1-2\rho_{\rm sat}$ and softening at $\gtrsim 4\rho_{\rm sat}$ are not allowed.
These characteristics can also be seen from the $\rho-c_{\rm s}$ plot in Fig.\ref{fig:posterior} where the relatively larger $c_{\rm s}$ at $\sim 1-2\rho_{\rm sat}$ and the relatively smaller $c_{\rm s}$ at $\sim 3-4\rho_{\rm sat}$ are not favored.
Meanwhile, as also found in other researches \citep{2020ApJ...893L..21R, 2020ApJ...892...55J, 2020PhRvD.101l3007L}, the inclusion of the data of {PSR J0030+0451} further improves the constraints.

For the bulk properties $R$ and $\Lambda$, the PSR mass information significantly increase the radius at low mass region, while the nuclear constraints have ruled out a good fraction of high value regions 
(see lower panels in Fig.\ref{fig:posterior}).
These larger $R$ and $\Lambda$ regions are further excluded by the GW data (see also the Tab.\ref{tab:posterior}).
While the data of {PSR J0030+0451} boost the lower bound of the $R$ and $\Lambda$ for $\sim 0.3~\rm km$ and $\sim 100$, respectively.
This can be explained by the relatively larger radius of {PSR J0030+0451} compared with the radii of NSs in GW170817 \citep{2018PhRvL.121p1101A}, albeit they have similar mass.

It has been conjectured that, in NS the speed of sound is less than the speed of light in vacuum divided by $\sqrt{3}$, i.e., $c_{\rm s}^2/c^2<1/3$, which is also called the conformal limit.
While the existence of the heavy NS ($\sim2~M_\odot$) may call for the violation of conformal limit \citep{2015PhRvL.114c1103B}.
Our results show that the 90\% lower limits of the $c_{\rm s}^2/c^2$ are larger than 1/3 in the high density region, which means that the conformal limit is indeed violated at some densities.

\begin{table}[]
    \caption{Credible intervals of Marginalized distributions of the properties for different sets of data. We give the median and 90\% credible intervals for the maximum mass of NS $M_{\rm max}$, radius for $1.4~M_\odot$ NS (i.e., $R_{\rm 1.4}$), radius for $2~M_\odot$ NS (i.e., $R_{\rm 2}$), tidal deformability for $1.4~M_\odot$ NS (i.e., $\Lambda_{1.4}$), and the pressure at $\rho_{\rm sat}$, $2\rho_{\rm sat}$, and $6\rho_{\rm sat}$, i.e., $p(\rho_{\rm sat})$, $p(2\rho_{\rm sat})$, and $p(6\rho_{\rm sat})$.}
    \begin{center}
        \begin{tabular}{l@{\quad}c@{\quad}c@{\quad}c@{\quad}c@{\quad}c}
            \hline \hline
            Observable & Prior                     & PSR                       & PSR+Nuclear             & PSR+Nuclear+GW          & PSR+Nuclear+GW+NICER    \\
            \hline
            \mmax      & $2.15^{+0.77}_{-0.68}$    & $2.53^{+0.42}_{-0.41}$    & $2.54^{+0.41}_{-0.43}$  & $2.37^{+0.51}_{-0.29}$  & $2.40^{+0.48}_{-0.32}$  \\
            \ronefour  & $18.60^{+2.38}_{-10.84}$  & $19.64^{+1.37}_{-9.49}$   & $12.66^{+1.51}_{-1.68}$ & $11.64^{+1.69}_{-1.48}$ & $11.83^{+1.25}_{-1.08}$ \\
            \rtwo      & $18.08^{+4.31}_{-10.71}$  & $20.12^{+2.36}_{-9.55}$   & $12.89^{+1.88}_{-2.17}$ & $11.66^{+1.69}_{-1.48}$ & $11.87^{+1.53}_{-1.43}$ \\
            \lonefour  & $5440^{+7131}_{-5421}$    & $7932^{+4819}_{-7771}$    & $518^{+680}_{-332}$     & $287^{+346}_{-152}$     & $323^{+334}_{-165}$     \\
            \onerho    & $17.58^{+15.90}_{-16.78}$ & $23.83^{+10.93}_{-22.98}$ & $4.41^{+0.91}_{-0.95}$  & $4.31^{+0.87}_{-0.91}$  & $4.27^{+0.93}_{-0.85}$  \\
            \tworho    & $4.17^{+6.96}_{-3.95}$    & $5.51^{+6.41}_{-5.18}$    & $4.64^{+7.29}_{-2.88}$  & $2.33^{+4.46}_{-0.74}$  & $2.59^{+4.37}_{-0.94}$  \\
            \sixrho    & $2.68^{+16.01}_{-2.25}$   & $4.27^{+17.30}_{-3.64}$   & $13.16^{+9.92}_{-8.62}$ & $12.51^{+9.13}_{-5.28}$ & $13.02^{+8.96}_{-5.92}$ \\
            \hline
            \hline
        \end{tabular}
    \end{center}
    \label{tab:posterior}
\end{table}

\section{Summary}
\label{summary}

In this work, we introduce a new nonparametric representation of the NS EoS via the FFNN.
Then we implement this nonparametric model to fit the theoretical EoSs including the hadronic EoSs, the hyperonic EoSs, and the quark matter (hybrid) EoSs.
We find that our method performs well in reconstructing these theoretical EoSs. For the absolute value of $\phi$ of the order $\sim 1 - 10$, the absolute fitting errors $|\phi - \phi_{\rm fit}|$ are on the order of $\sim 10^{-4} - 10^{-1}$.
And the relative errors $|1-\Gamma_{\rm fit}/\Gamma|$ in calculating the corresponding $\Gamma(p)$ are as low as $\sim$ 0.1\% $-$ 10\%.
Then we make injections and recover them using the same procedure as what we applied on the real data analysis.
We inject mock observations of three NSs considering the masses of GW170817 and PSR J0030+0451, and the tidal deformabilities and radius of these NSs are calculated by four theoretical EoSs, i.e., WFF1, H4, MS1, and ALF2.
We find that there are no egregious errors on recovering both the macroscopic properties and microscopic relations with simulated data.
After validating our model, we further adopt it to analyze the real observation data, i.e., mass-deformability measurement of GW170817 and mass-radius measurement of {PSR J0030+0451}, and use Bayesian method to constrain the EoS parameters.
We find that the radius and deformability of a canonical $1.4~M_\odot$ neutron star are correspondingly $R_{1.4}=11.83^{+1.25}_{-1.08}~{\rm km}$ and $\Lambda_{1.4} = 323^{+334}_{-165}$ (see the bottom panels of Fig.\ref{fig:posterior}), where the credible intervals are all 90\%.
These results are in agreement with those found in the previous investigations \citep{2018A&A...616A.105S, 2019ApJ...887L..22R, 2019ApJ...887L..24M, 2020ApJ...892...55J, 2020ApJ...893L..21R, 2020PhRvD.101l3007L, 2020arXiv200203210Z, 2020arXiv200801582B, 2021PhRvC.103b5803H, 2021PhRvL.126f1101A}.

The speed of sound in NS is another significant property of the EoS.
There is a conjecture that the sound speed in the NS satisfies the so called conformal limit, i.e., $c_{\rm s}^2/c^2<1/3$. Using our nonparametric method, we infer that the square of the sound speed in the core of the maximum mass configuration NS is larger than $c^2/3$ at 90\% credible level (see the top right panel of Fig.\ref{fig:posterior}), that means the conformal limit is violated in the center of very massive NSs.
\citet{2019ApJ...887L..22R} have imposed a constraint that the speed of sound of each EoS converges to conformal limit at asymptotically high densities according to pQCD calculations \citep{2014ApJ...781L..25F}, in their speed of sound (CS) model.
While our model does not guarantee this constraint, but notice that the central densities of the observed NSs are far away from pQCD density, this constraint will not significantly affect our results.

As mentioned in Sec.\ref{nonp method}, the finite sum of Eq.(\ref{eq:FFNN}) can fit the continuous function at high resolution, and the larger the number of the nodes is the better the approximation gets.
Therefore, we need to extend the width of the hidden layer(increase the number of the neural nodes) to get better approximation in the future.
However, larger neural networks have more parameters, which may be extremely hard for the sampling algorithm to converge.
Thus a `smarter' sampling algorithm like {\sc PolyChord} \citep{2015MNRAS.453.4384H} that can support higher dimensional parameter space would be considered, and the balance between the performance of fitting and the computational cost would also be carefully considered.
Besides, more complex model would lead to the over-fitting problem (i.e., the model can better fit the current data, but worse predict the unknown data point.), and the posterior inference of more complex models may dramatically change as more data is accrued. Therefore, in future works we should pay more attention to the trade-off between the variance and bias to get a better model.

\section{Acknowledgments}
We thank Mr. Lei Zu for discussing the neural networks.
This work was supported in part by NSFC under grants of No. 11921003 and No. 11525313. This research has made use of data, software and/or web tools obtained from the Gravitational Wave Open Science Center (\url{https://www.gw-openscience.org/}), a service of LIGO Laboratory, the LIGO Scientific Collaboration and the Virgo Collaboration. LIGO Laboratory and Advanced LIGO are funded by the United States National Science Foundation (NSF) as well as the Science and Technology Facilities Council (STFC) of the United Kingdom, the Max-Planck-Society (MPS), and the State of Niedersachsen/Germany for support of the construction of Advanced LIGO and construction and operation of the GEO600 detector. Additional support for Advanced LIGO was provided by the Australian Research Council. Virgo is funded, through the European Gravitational Observatory (EGO), by the French Centre National de Recherche Scientifique (CNRS), the Italian Istituto Nazionale di Fisica Nucleare (INFN) and the Dutch Nikhef, with contributions by institutions from Belgium, Germany, Greece, Hungary, Ireland, Japan, Monaco, Poland, Portugal, Spain.
\bibliography{bibtex}

\end{document}